\documentclass[aps,prd,nofootinbib,amsmath,twocolumn]
{revtex4-1}
\usepackage{amssymb,esvect,amsmath,graphicx,latexsym,amsthm,slashed,eso-pic}
\usepackage[final]{pdfpages}

\newcommand{\beq}{\begin{equation}} \newcommand{\eeq}{\end{equation}}
\newcommand{\bea}{\begin{eqnarray}} \newcommand{\eea}{\end{eqnarray}}

\newcommand{\be}{\begin{eqnarray}}
\newcommand{\ee}{\end{eqnarray}}

\newcommand{\zp}{\, Z^\prime}

\def\lsim{\mathrel{\raise.3ex\hbox{$<$\kern-.75em\lower1ex\hbox{$\sim$}}}}
\def\gsim{\mathrel{\raise.3ex\hbox{$>$\kern-.75em\lower1ex\hbox{$\sim$}}}}

\begin{document}

\hspace{13cm} \parbox{5cm}{FERMILAB-PUB-16-040-A}~\\



\title{PeV-Scale Dark Matter as a Thermal Relic of a Decoupled Sector}
\author{Asher Berlin,$^1$ Dan Hooper,$^{2,3}$ and Gordan Krnjaic$^2$}

\affiliation{$^1$ Department of Physics, Enrico Fermi Institute, University of Chicago, Chicago, IL}
\affiliation{$^2 $Center for Particle Astrophysics, Fermi National Accelerator Laboratory, Batavia, IL  60510}
\affiliation{$^3$ Department of Astronomy and Astrophysics, The University of Chicago, Chicago, IL  60637}

\date{\today}

\begin{abstract}

In this letter, we consider a class of scenarios in which the dark matter is part of a heavy hidden sector that is thermally decoupled from the Standard Model in the early universe. The dark matter freezes-out by annihilating to a lighter, metastable state, whose subsequent abundance can naturally come to dominate the energy density of the universe. When this state decays, it reheats the visible sector and dilutes all relic abundances, thereby allowing the dark matter to be orders of magnitude heavier than the weak scale. For concreteness, we consider a simple realization with a Dirac fermion dark matter candidate coupled to a massive gauge boson that decays to the Standard Model through its kinetic mixing with hypercharge. We identify viable parameter space in which the dark matter  can be as heavy as $\sim$1-100 PeV without being overproduced in the early universe. 


\end{abstract}

\maketitle
The Weakly Interacting Massive Particle (WIMP) paradigm provides a compelling cosmological origin for dark matter (DM) candidates with weak-scale masses and interactions.
In the early universe, at temperatures above the WIMP's mass, interactions with the Standard Model (SM) produce a thermal population of WIMPs and sustain chemical equilibrium between dark and visible matter. When the temperature falls below the WIMP's mass, these interactions freeze-out to yield an abundance similar to the observed cosmological DM density. This narrative is known as the ``WIMP miracle."

In recent years, however, this framework has become increasingly constrained. The Large Hadron Collider has not yet discovered any new physics, and limits from direct detection experiments have improved at an exponential rate over the past decade. For DM candidates that annihilate at a sufficient rate to avoid being overproduced in the early universe, unacceptably large elastic scattering cross sections with nuclei are often predicted. To evade these constraints, one is forced to consider models that include features such as coannihilations~\cite{Griest:1990kh,Edsjo:1997bg}, resonant annihilations~\cite{Griest:1990kh,Hooper:2013qjx}, pseudoscalar couplings~\cite{Boehm:2014hva,Ipek:2014gua,Berlin:2014tja,Berlin:2015wwa}, or annihilations to final states consisting of leptons or electroweak bosons~\cite{Drees:1992rr,Hisano:2010ct,Hisano:2011cs,Hisano:2015bma,Ibarra:2015fqa,Hill:2014yka,Hill:2014yxa,Agrawal:2011ze,Berlin:2015ymu}.

It is equally plausible, however, that the DM is a singlet under the SM and was produced independently of the visible sector during the period of reheating that followed inflation (for a review, see Ref.~\cite{Allahverdi:2010xz}). By freezing-out through annihilations to SM singlets, the DM in such models can avoid being overproduced while easily evading the constraints from direct detection experiments~\cite{Pospelov:2007mp,Pospelov:2008jd,ArkaniHamed:2008qn,Cholis:2008qq,Morrissey:2009ur,Meade:2009rb,Andreas:2011in,Hooper:2012cw}. In this letter, we explore this class of scenarios, focusing on hidden sectors that are thermally decoupled and, therefore, never reach equilibrium with the visible sector. In this case, the DM freezes-out of chemical equilibrium within its own sector, unaffected by SM dynamics.

So long as the hidden sector consists entirely of SM singlets, renormalizable interactions between the SM and the DM can proceed only through the following gauge singlet operators: $H^\dagger H$, $B^{\mu\nu}$, and $H^\dagger L$, known as the Higgs portal~\cite{Burgess:2000yq,Pospelov:2007mp,Davoudiasl:2004be,Bird:2006jd,Kim:2006af,Finkbeiner:2007kk,D'Eramo:2007ga,Barger:2007im,SungCheon:2007nw,MarchRussell:2008yu,McDonald:2008up,Pospelov:2011yp,Piazza:2010ye,Kouvaris:2014uoa,Kainulainen:2015sva,Krnjaic:2015mbs}, the vector portal ~\cite{Pospelov:2007mp,Krolikowski:2008qa}, and the lepton portal~\cite{Pospelov:2007mp,Bai:2014osa}, respectively.
%
If the couplings that facilitate such interactions are sufficiently small, the hidden and visible sectors will be decoupled from one another, potentially altering the thermal history of the universe~(see, e.g., Refs.~\cite{Feng:2008mu,Cheung:2010gj,Sigurdson:2009uz,Kuflik:2015isi,Pappadopulo:2016pkp}).

If, by coincidence, a hidden sector DM candidate has a GeV-TeV scale mass and weak-scale couplings, it will behave in many respects like a typical WIMP, although possibly with very feeble interactions with the SM.
Alternatively, if the hidden sector is much heavier than the SM, its lightest particles may be long-lived and come to dominate the energy density of the universe. When these states ultimately decay through portal interactions, they can deposit significant entropy into the SM bath, thereby diluting the naively excessive DM abundance. Thus, in this class of models, the DM may be much heavier than the mass range typically favored by standard thermal relic arguments; as large as $\sim$1-100 PeV without exceeding the measured cosmological dark matter density.

Although the mechanism described in this letter could be realized within the context of the Higgs, vector, or lepton portals, for concreteness we will focus here on the vector portal scenario. For our DM candidate, we introduce a stable Dirac fermion, $X$, which has unit charge under a spontaneously broken $U(1)_X$ gauge symmetry, corresponding to the massive gauge boson, $\zp$. The hidden Lagrangian contains: 
\be
{\cal L}&\supset& -\frac{\epsilon}{2} B^{\mu\nu}\zp_{\mu \nu} +   g_{\rm DM} Z^{\prime}_{\mu}  \overline X \gamma^\mu X,~~ 
\ee
where $\zp_{\mu \nu}$ and $B_{\mu\nu}$ are the $U(1)_X$ and hypercharge field strengths, respectively, and $\epsilon$ quantifies their kinetic mixing~\cite{Holdom:1985ag,Okun:1982xi}. A  small, non-zero value of  $\epsilon$ can be radiatively generated if heavy $U(1)_{X} \times U(1)_Y$ charged particles are integrated out at some high scale. Since any value of $\epsilon$ is technically natural, it is generic to expect  $\epsilon \ll 1$. Thus, if $Z^\prime$ is the lightest hidden sector particle, it can easily be very long-lived, leading it to dominate the energy density of the universe and change significantly the predictions of thermal freeze-out.

The thermal freeze-out from chemical equilibrium of the $X$ population is dictated by their annihilation cross section which, for $m_X > m_{Z^\prime}$, is given by:
\be  \label{eq:approx-annihilation-rate}
\sigma v_{ X\bar{X} \to Z' Z'} \simeq \frac{ \pi \alpha^2_X}{ m_X^2 }\,,~ 
\ee
where $\alpha_X \equiv g_{\rm DM}^2/4\pi$ and we have dropped subleading terms (see Supplementary Material, Sec.~\ref{sigmav}).\footnote{In the $m_{\zp} >  m_X$ regime, the dominant annihilation channel is $X\bar X \to {\zp} \to$ SM, with a cross section that is proportional to $\epsilon^2$. If $\zp$ is long-lived, this annihilation cross section will be too small to facilitate a viable thermal freeze-out~\cite{Izaguirre:2015yja}.}  This leads to a relic abundance comparable to the measured dark matter density for weak-scale couplings and masses, $\alpha_X \sim 0.0035 \times (m_X/100 \, {\rm GeV})$.    
%
%
Although somewhat heavier DM particles with larger couplings are also possible, partial-wave unitarity 
imposes a constraint on $\alpha_X$, which translates into a hard upper limit of $m_X \lsim$100 TeV~\cite{Griest:1989wd}. This bound can be comfortably circumvented, however, if the hidden and visible sectors are decoupled at early times. 

As an initial condition, we take the hidden and visible sectors to be described by separate thermal distributions, with temperatures of $T_h$ and $T$, respectively. The ratio of these temperatures, $\xi_{\rm inf} \equiv (T_h/T)_{\rm inf}$, is determined by the physics of inflation, including the sectors' respective couplings to the inflaton~\cite{Hodges:1993yb,Berezhiani:1995am}. Using entropy conservation in each sector, we can calculate the time evolution of $\xi$ (prior to the decays of $Z'$):
\begin{eqnarray}
\frac{s_h}{s} &=& \frac{g^h_{\star}}{g_{\star}} \, \xi^3 = {\rm constant} \\
&\rightarrow& \xi = \xi_{\rm inf} \, \bigg(\frac{g^h_{\star, {\rm inf}}}{g^h_{\star}}\bigg)^{1/3} \, \bigg(\frac{g_{\star}}{g_{\star,{\rm inf}}}\bigg)^{1/3}, \nonumber
\end{eqnarray}
where $g_{\star}$ and $g_{\star}^h$ are the numbers of effective relativistic degrees-of-freedom in the visible and hidden sectors, respectively. If the SM temperature is well above the electroweak scale, $g_{\star} \simeq g_{\star,{\rm inf}}$. As the temperature of the hidden sector falls below $m_X$, $g^h_{\star}$ decreases from $g_{Z'}+(7/8)g_X$ to $g_{Z'}$, bringing $\xi$ from $\xi_{\rm inf}$ to $(13/6)^{1/3} \, \xi_{\rm inf} \approx 1.3 \, \xi_{\rm inf}$, for $m_{Z^\prime}\ll m_X$.



As the universe expands, $X$ will eventually freeze-out of chemical equilibrium, yielding a non-negligible relic abundance. The evolution of the number density of $X$ (plus $\bar{X}$), $n_X$, is described by the Boltzmann equation:
\be
\label{eq:Boltz1}
\dot{n}_X + 3 H  n_X = - \frac{1}{2}\langle \sigma v \rangle ~ \left(n_X^2 - \frac{n^2_{Z'}}{\, n^2_{Z', {\text{eq}}}}n_{X,{\text{eq}}}^2 \right),
\ee
where $\langle \sigma v \rangle$ is the thermally averaged cross section for the process $X\, \bar{X} \rightarrow Z' \, Z'$, $H = [8 \pi \, (\rho_{\rm SM} + \rho_h)/3 \, m^2_{\rm Pl}]^{1/2}$ describes the expansion rate of the universe in terms of the energy densities in the visible and hidden sectors, and $m_{\rm Pl} \simeq 1.22 \times 10^{19}$ GeV. Here, we have assumed that $n_X = n_{\bar{X}}$. Note that this expression allows for the possibility that the $Z'$ number density is not equal to the equilibrium value, as the $Z'$ population is also expected to freeze-out of equilibrium during this epoch.

\begin{figure}[t]
\hspace{-0.5cm}\includegraphics[width=0.51\textwidth]{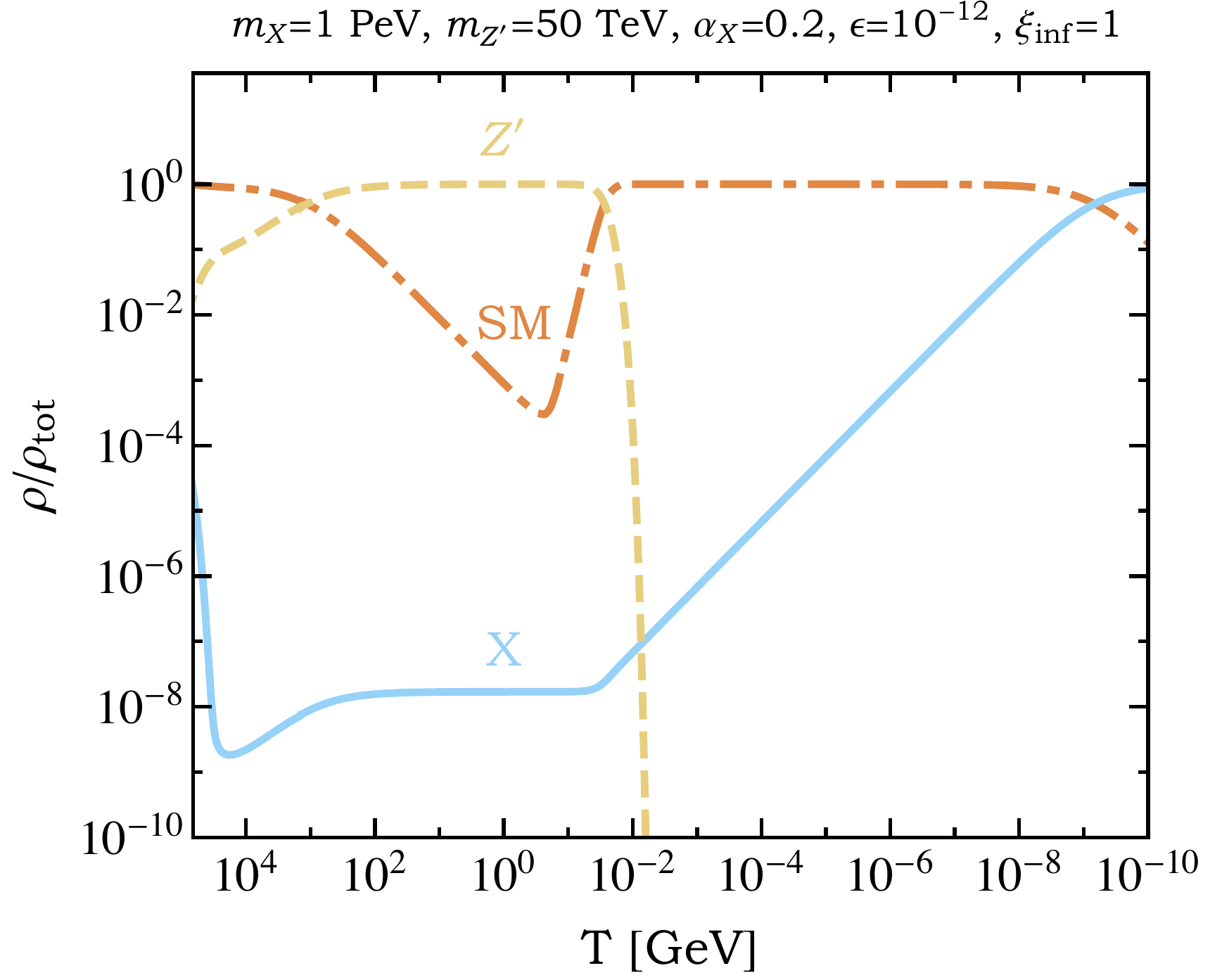}
\caption{The evolution of the energy densities of dark matter (blue solid), of $Z'$s (yellow dashed), and in the visible sector (orange dot-dashed), as a function of the visible sector temperature. Upon becoming non-relativistic, the $Z'$s quickly come to dominate the energy density of the universe and, when they decay, they heat the SM bath and dilute the $X$ abundance. This is a rather generic feature of models with a heavy and decoupled hidden sector.}
\label{example}
\end{figure}

In the case that 
$n_{Z'}$ remains close to its equilibrium value during the freeze-out of $X$ (see Supplementary Material, Sec.~\ref{zprimefo}), the Boltzmann equation can be solved semi-analytically. In this case, the thermal relic abundance of $X$ (plus $\bar{X}$) is given by:
\begin{eqnarray}
\label{relic1}
  \hspace{-0.1cm} \Omega_X h^2 &\approx& 8.5 \times 10^{-11} ~ \frac{x_f \sqrt{g_\star^\text{eff}}}{g_*} ~ \left( \frac{a+3 \xi b / x_f}{\text{GeV}^{-2}} \right)^{-1} ~~ \\
   && \hspace{-0.3cm} \approx   1.6 \times 10^4 \bigg(\frac{x_f}{30}\bigg)  \bigg(\frac{0.1}{\alpha_X}\bigg)^2  \bigg(\frac{m_X}{{\rm PeV}}\bigg)^2
\bigg(\frac{\sqrt{g^{\rm eff}_{\star}}/g_{\star}}{0.1} \!  \bigg),~~~~ \label{relic2} \nonumber
\end{eqnarray}
where $a$ and $b$ are terms in the expansion of the DM annihilation cross section, $\sigma v/2 \approx a +b v^2 + \mathcal{O}(v^4)$ (see Supplementary Material, Sec.~\ref{sigmav}), and $g^{\rm eff}_{\star} \equiv g_{\star} + g^h_{\star} \, \xi^4$ at freeze-out. $x_f$, which is defined as the mass of $X$ divided by the SM temperature at freeze-out, is found to be $\sim 20 \times \xi$ over a wide range of parameters (see Supplementary Material, Sec.~\ref{fotemp}).
From Eq.~\ref{relic1}, it is clear that a PeV-scale DM candidate with perturbative couplings will initially freeze-out with an abundance that exceeds the observed DM density ($\Omega_X h^2 \gg \Omega_{\rm DM} h^2 \simeq 0.12$). 
It has long been appreciated, however, that this conclusion can be circumvented if the universe departed from the standard radiation-dominated picture after DM freeze-out~\cite{Fornengo:2002db,Gelmini:2006pq,Kane:2015jia,Hooper:2013nia,Patwardhan:2015kga,Randall:2015xza,Patwardhan:2015kga,Reece:2015lch,Lyth:1995ka,Davoudiasl:2015vba,Cohen:2008nb,Boeckel:2011yj,Boeckel:2009ej,Hong:2015oqa}. A novel point that we emphasize here is that such a departure is generically expected within the context of hidden sector models with small couplings to the visible sector.  More specifically, as the universe expands, the remaining $Z'$s will become non-relativistic and quickly come to dominate the energy density of the universe when  $\rho_{Z'} =  0.0074 \, g_* \xi_{\rm inf}^3  \,  m_{Z'} T_{\rm dom}^3 >  (\pi^2/30)g_* T_{\rm dom}^4$, which occurs at a visible sector temperature of:
 \be
  T_{\rm dom}  
  \sim 1 \, {\rm TeV} \, \times \, \xi_{\rm inf}^3 \left( \frac{m_{Z^\prime}}{50 \, \rm TeV} \right).~~
   \ee
This expression is valid so long as the $Z'$s depart from chemical equilibrium while relativistic.  
 When the $\zp$ population ultimately decays, it will deposit energy and entropy into the visible sector, potentially diluting the DM abundance to acceptable levels. In Fig.~\ref{example}, we show the evolution of the energy densities in the visible and hidden sectors, for a representative choice of parameters in this model.

\begin{figure*}[t]
\hspace{-2.cm}\includegraphics[width=0.51\textwidth]{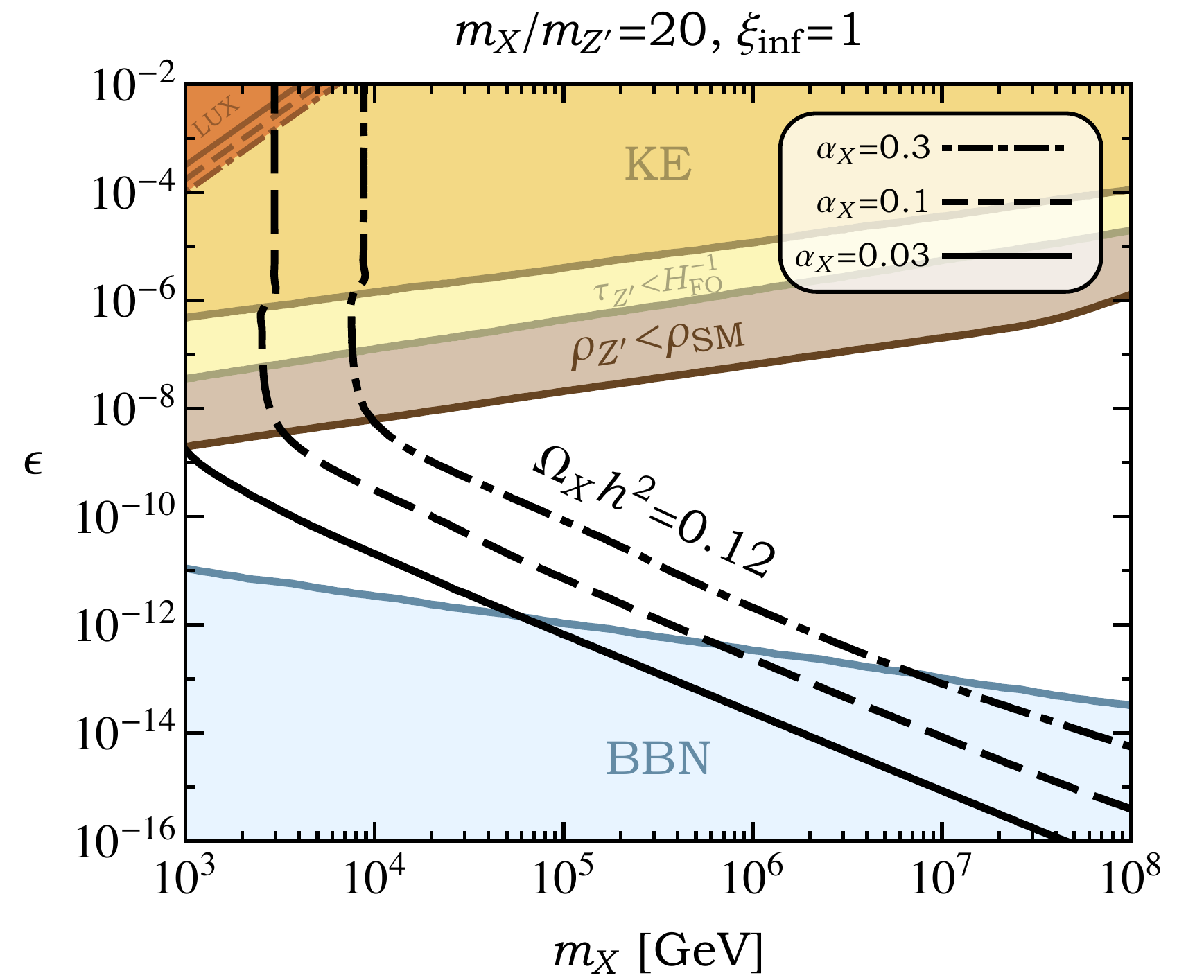} \hspace{-0.5cm}
\includegraphics[width=0.51\textwidth]{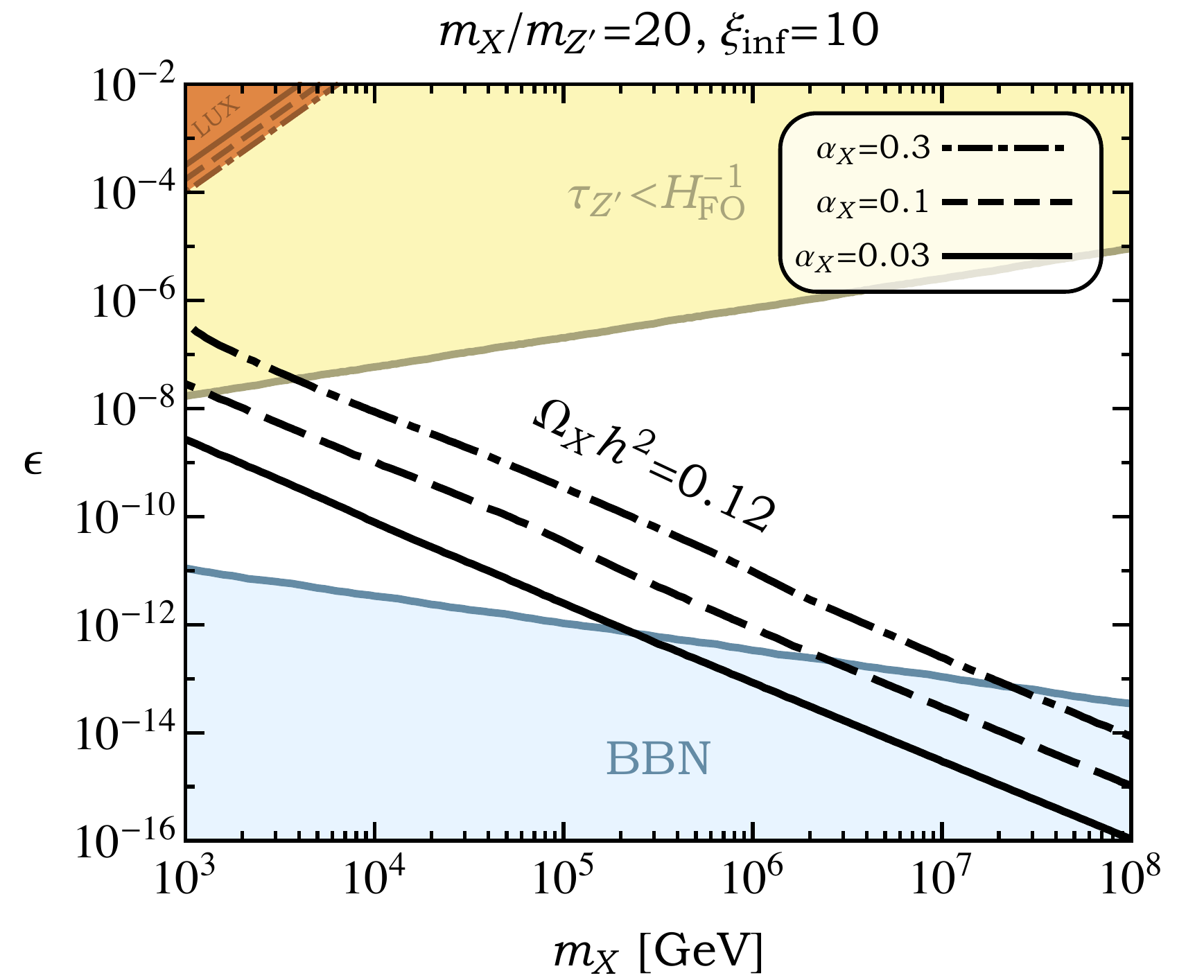} \hspace{-2cm}
\caption{The black contours represent the regions of the $m_X-\epsilon$ plane in which the dark matter density is equal to the measured cosmological abundance, for three values of the hidden sector interaction strength, $\alpha_X$, and for $m_{Z'} = m_{X}/20$. In the left panel, we adopt equal initial temperatures for the hidden and visible sectors, $\xi_{\rm inf} \equiv (T_{h}/T)_{\rm inf}=1$. In the right panel, we instead assume that the universe was highly dominated by the hidden sector after inflation, $\xi_{\rm inf}=10$. In each panel, the red and blue regions are excluded by direct detection and BBN constraints, respectively. In and above the orange and yellow regions, the hidden and visible sectors are in kinetic equilibrium during dark matter freeze-out, or the $Z'$ population decays before the freeze-out of $X$, respectively. In and above the brown region, the $Z'$ population never dominates the energy density of the universe, and thus does not significantly dilute the dark matter relic abundance. In contrast to the case of a standard thermal relic, dark matter from a decoupled sector can be as heavy as $\sim$1-100 PeV without being overproduced in the early universe.}
\label{mainfig}
\end{figure*}

For a simple estimate of this effect, suppose that all of the $Z'$s decay at $t=\tau_{Z'}$. Immediately prior to their decays, they dominate the energy density as non-relativistic matter, so $H=2/3 \tau_{Z'}$. Combining this with the Friedmann equation, $H^2 = 8\pi \rho_{Z'}/3m^2_{\rm Pl}$, we find:
\begin{eqnarray}
\label{eq:Hdecay}
\frac{4}{9 \, \tau^2_{Z'}} 
\approx \, 0.062 \, g_* \, \xi_\text{inf}^3 \, \frac{m_{Z'} T_i^3}{m_\text{Pl}^2} .
\end{eqnarray}
Thus the temperature of the visible sector immediately prior to the decays is given by:
\begin{eqnarray}
 T_i \approx  \frac{  0.31 \,{\rm GeV}  }{\xi_{\rm inf}} \bigg(\frac{\epsilon}{10^{-10}}\bigg)^{4/3} \bigg(\frac{m_{Z'}}{100\,{\rm TeV}}\bigg)^{1/3} \bigg(\frac{100}{g_{\star}}\bigg)^{1/3} \! .~~~~
\end{eqnarray}
%
 From energy conservation ($\rho_\text{SM} = \rho_{Z'}$), the temperature of the SM bath immediately following the $Z^\prime$ decays is
 set by the relation $(\pi^2/30) g_{\star} T_f^4 = m^2_{\rm Pl}/6 \pi \tau^2_{Z'}$,
which yields the final visible sector temperature:  
\begin{eqnarray}
 T_f  \approx  \, {\rm GeV} \, \bigg(\frac{\epsilon}{10^{-10}}\bigg) \, \bigg(\frac{m_{Z'}}{100 \, {\rm TeV}}\bigg)^{1/2} \, \bigg(\frac{100}{g_{\star}}\bigg)^{1/4}\!\!.~~~~ 
\end{eqnarray}
As a consequence of the reheating that results from these decays, the abundances of any previously frozen-out relics (including $X$) will be diluted by a factor of $(T_f/T_i)^3$:
\be
\frac{S_f}{S_i} \sim 800 \times \left( \frac{10^{-10}}{\epsilon}\right) \left(\frac{m_{Z'}}{100\,{\rm TeV}}\right)^{1/2} \, \left( \frac{g_*}{100} \right)^{1/4} \, \xi^3_{\rm inf}  ~.~~~~~~
\ee
A more careful calculation, integrating over the $Z'$ decay rate \cite{Kolb:1990vq}, yields:
\begin{eqnarray}
~ \frac{S_f}{S_i} \approx 680 \times \bigg( \!\frac{10^{-10}}{\epsilon} \!  \bigg) \bigg(\frac{m_{Z'}}{100\,{\rm TeV}}\bigg)^{1/2} \bigg(  \frac{\langle g^{1/3}_{\star} \rangle^{3}}{100}   \bigg)^{1/4} \, \xi_{\rm inf}^3,~~~~~
\end{eqnarray}
where $\langle g_{\star} \rangle$ denotes the time-averaged value over the period of decay.
Combining this with Eq.~\ref{relic1}, we find that the final DM relic abundance is:
\begin{eqnarray}
~\Omega_X h^2 &\approx & \frac{0.12}{\xi_{\rm inf}^3} \,         \bigg(\frac{\epsilon}{10^{-13}}\bigg) \, \bigg(\frac{0.045}{\alpha_X}\bigg)^2 \, \bigg(\frac{m_X}{{\rm PeV}}\bigg)^2 \, \bigg(\frac{100 \,{\rm TeV}}{m_{Z'}}\bigg)^{1/2}   \nonumber  \\
 &&  \hspace{1 cm} \times    \bigg(\frac{x_f}{30}\bigg)    \bigg(\frac{\sqrt{g^{\rm eff}_{\star}}/g_{\star}}{0.1}\bigg) \,
  \bigg(  \frac{100}{\langle g^{1/3}_{\star} \rangle^{3}} \bigg)^{1/4} . \!\!\!\!\!\!
\end{eqnarray}

In Fig.~\ref{mainfig} we plot some of the phenomenological features of this model as a function of the DM mass and the degree of kinetic mixing between the $Z'$ and SM hypercharge. The black contours denote the regions where the DM density is equal to the measured cosmological abundance, for three values of the hidden sector interaction strength, $\alpha_X$. Below the brown region, $Z'$ decays deposit significant entropy into the visible sector, reducing the final $X$ abundance.

Also plotted in this figure are the constraints from the null results of direct detection experiments and the successful predictions of Big Bang Nucleosynthesis (BBN).  Comparing the elastic scattering cross section between DM and nuclei predicted in this model to the most recent constraints from LUX~\cite{Akerib:2015rjg} (for a value of $\alpha_X$ that yields the desired thermal relic abundance, again assuming that $m_X > m_{Z'}$), we arrive at a constraint of $\epsilon \lsim 1.1 \times 10^{-3} \times (m_{Z'}/100 \, {\rm GeV})^2$, for $m_X \gsim 50$ GeV. To assure consistency with BBN, we require that the temperature of the universe exceeds 10 MeV after the decays of the $Z'$ population, resulting in the following constraint: $\epsilon \gsim 2 \times 10^{-13} \times (100 \, {\rm TeV}/m_{Z'})^{1/2} \, (g_{\star}/10)^{1/4}$ (see Supplementary Material, Secs.~\ref{direct}-\ref{bbnsec}). 

The constraints described in the previous paragraph can be satisfied for a wide range of $\epsilon$, spanning many orders of magnitude. Depending on the degree of kinetic mixing, the hidden and visible sectors may have been entirely decoupled from one another, or kept in kinetic equilibrium through interactions of the type $\gamma f \leftrightarrow Z' f$ (see Supplementary Material, Sec.~\ref{kesec}).
Quantitatively, we find that the rate for these processes exceed that of Hubble expansion if: $\epsilon \gsim  10^{-7} \times (T/10\,{\rm GeV})^{1/2}$ (shown as the orange region in Fig.~\ref{mainfig}).  Thus for smaller values of $\epsilon$, the hidden sector will not reach equilibrium with the visible sector and will remain decoupled. Furthermore, in the yellow regions of Fig.~\ref{mainfig}, the $Z'$ population decays prior to the freeze-out of $X$.

In Fig.~\ref{example} and in the left-panel of Fig.~\ref{mainfig}, we have presented results for a case in which the visible and hidden sectors were initially reheated to similar temperatures after inflation, $\xi_{\rm inf} = 1$. It is also interesting to consider scenarios in which the initial temperatures of these sectors are very different. In the $\xi_{\rm inf} \ll 1$ case, the $Z'$ population does not come to dominate the energy density of the universe, and their decays do not significantly impact cosmological history. The DM in this scenario, however, is produced with a relic abundance that is proportional to $\xi$, making it possible to avoid overproduction even for very large masses. An even more interesting case is that in which reheating preferentially populates the hidden sector, with comparatively little SM particle content ($\xi_{\rm inf} \gg 1$), corresponding to the right-panel of Fig.~\ref{mainfig}. In this case, the energy density of the universe will remain dominated by the hidden sector until the $Z'$ population decays, thereby generating the SM bath. In the $\xi_{\rm inf} \gg 1$ limit, the final abundance of DM is approximately given by:
%
\begin{eqnarray}
\hspace{-0.1cm}\Omega_X h^2 
&\sim& 0.12 \times  \bigg(\frac{0.06}{\alpha_X}\bigg)^2 \, \bigg(\frac{m_X}{{\rm PeV}}\bigg)^2 \, \bigg(\frac{100\,{\rm TeV}}{m_{Z'}}\bigg)^{1/2} \bigg( \frac{100}{g_*}\bigg) \,\nonumber \\
&& ~~~~~\times  \bigg(  \frac{100}{\langle g^{1/3}_{\star} \rangle^{3}} \bigg)^{1/4} \, \bigg(\frac{\epsilon}{10^{-12}}\bigg)   \, \bigg(\frac{\xi/\xi_{\rm inf}}{1.3}\bigg)^{3}. 
\label{relic}
\end{eqnarray}
This allows for an acceptable $X$ abundance, without violating the constraints from BBN, for masses as high as:
\begin{eqnarray}
m_X \lsim 40 \, {\rm PeV} \, \bigg(\frac{\alpha_X}{0.3}\bigg)^2 \, \bigg(\frac{10}{m_{X}/m_{Z'}}\bigg),~
\end{eqnarray}
where we have taken $g_* \approx 10$ near BBN temperatures. 
If we select a value of $\alpha_X$ that saturates the unitarity bound~\cite{Griest:1989wd}, this scenario allows for DM as heavy as $m_{X} \sim 5 \, {\rm EeV} \times [10/(m_X/m_{Z'})]$.



                    


In this letter, we have considered a class of scenarios in which the DM resides within a heavy sector that is highly decoupled from the Standard Model. When the temperature falls below the mass of the lightest hidden sector particle, this long-lived state is expected to rapidly come to dominate the energy density of the universe, ultimately heating the visible sector and diluting the DM abundance through its decay. In contrast to conventional WIMPs, DM candidates as heavy as $\sim$1-100 PeV can be thermal relics of a decoupled hidden sector, without being overproduced in the early universe.  Although we have focused on a particular vector portal model in this letter, 
%
%
we emphasize that similar phenomenology can appear within the context of other DM models with a heavy hidden sector. 


\section*{Acknowledgments}
AB is supported by the Kavli Institute for Cosmological Physics at the University of Chicago through grant NSF PHY-1125897. Fermilab is operated by Fermi Research Alliance, LLC, under Contract No. DE-AC02-07CH11359 with the US Department of Energy. 

\newpage

\begin{appendix}

\section{Supplementary Material}

\subsection{Dark Matter Annihilation}
\label{sigmav}

In this model, the DM annihilation cross section can be written as an expansion in powers of velocity:
\begin{eqnarray}
\frac{1}{2} \sigma_{X\bar{X} \rightarrow Z' Z'}v \simeq a+bv^2 + \mathcal{O}(v^4),
\end{eqnarray}
where the s-wave piece is:
\be
a&=&\frac{2\pi \alpha_X^2}{m^2_{X}} \, \frac{(1-r^2)^{3/2}}{(2-r^2)^2}, 
\ee
 the p-wave contribution is: 
\be
b&=& \, \frac{    \pi \alpha_X^2   (1-r^2)^{1/2} \, (24+28r^2-36r^4+17r^6)    }{ 12 m^2_{X}  (2-r^2)^4}, ~~\nonumber
\ee
and we define $r \equiv m_{Z'}/m_X$.

\subsection{$Z'$ Couplings To Standard Model Fermions}

The $Z'$ couples to SM fermions through kinetic mixing with hypercharge. Following Ref.~\cite{Hoenig:2014dsa}, these vector and axial couplings are given by $g_{fv,fa} \equiv (g_{f_R}\pm g_{f_L})/2$, where
\begin{eqnarray}
g_{f_{R,L}} \! = \! \epsilon \bigg( \frac{m^2_{Z'} \, g_Y \, Y_{f_{R,L}} - m^2_Z \, g \,  \sin \theta_W \cos \theta_W \, Q_{f}}{m^2_{Z}-m^2_{Z'}}\biggr).~~~~~~
\end{eqnarray}
Here, $\theta_W$ is the weak mixing angle, $m_Z$ is the $Z$ mass as predicted in the SM,
and $g_Y$ and $g$ are the $U(1)_Y$ and $SU(2)_L$ gauge couplings, respectively.

\subsection{$Z'$ Freeze-Out}
\label{zprimefo}

At very high temperatures, $T_h \gg m_{Z'}, m_{X}$, a number of interactions will be able to maintain equilibrium among the particles in the hidden sector. As $T_h$ drops below $m_{Z'}$ and/or $m_{X}$, however, such processes become suppressed, ultimately leading to the freeze-out of the comoving $Z'$ number density. In this section, we estimate the temperature at which this freeze-out occurs.

We first consider interactions of the type $Z' Z' Z'  \rightarrow Z' Z'$, mediated by a $X$ loop. In analogy with the procedure followed in Ref.~\cite{Hochberg:2014dra}, dimensional analysis suggests that this corresponds to an operator of the form $\alpha^{5/2}_X F^5_{Z'} /m^6_X$, where $F_{Z'}$ is the $Z'$ field strength. The rate for such interactions can thus be estimated by:
\begin{eqnarray}
\Gamma_{Z' Z' Z'  \rightarrow Z' Z'} &=&  n^2_{Z'} \langle \sigma v^2 \rangle 
= n^2_{Z'} \frac{\Delta_1 \alpha^5_X T^7_h}{m^{12}_X}, 
\end{eqnarray}
where $\Delta_1$ is an order one (or smaller) coefficient intended to parameterize our ignorance of the cross section. In the $T_h \gg m_{Z'}$ limit, this scattering rate exceeds the rate of Hubble expansion when the following condition is met:
\begin{eqnarray}
\bigg(\frac{\zeta(3)g_{Z'} T^3_h}{\pi^2}\bigg)^2 \, \frac{\Delta_1 \alpha^5_X T^7_h}{m^{12}_X} \gsim \bigg(\frac{4\pi^3 g^{\rm eff}_{\star} T^4_h}{45 m^2_{\rm Pl} \xi^4}\bigg)^{1/2},
\end{eqnarray}
which reduces to:
\begin{eqnarray}
\frac{T_h}{m_X} \gsim \frac{0.3}{\Delta_1^{1/11}} \,   \bigg(\frac{1.3}{\xi}\bigg)^{2/11}  \bigg(\frac{m_X}{{\rm PeV}}\bigg)^{1/11} \bigg(\frac{0.1}{\alpha_X}\bigg)^{5/11}  \bigg(\frac{g^{\rm eff}_{\star}}{100}\bigg)^{1/22}. \nonumber \\
\end{eqnarray}

Next, we consider processes of the type $Z' XX \rightarrow XX$, 
\begin{eqnarray}
\Gamma_{Z' XX \rightarrow XX} &=&  n^2_{X} \langle \sigma v^2 \rangle \\
&=& n^2_{X} \frac{\Delta_2 \alpha^3_X}{m^{5}_X},\nonumber
\end{eqnarray}
and $Z' Z' X \rightarrow Z' X$:
\begin{eqnarray}
\Gamma_{Z' Z' X \rightarrow Z' X} &=&  n_{X} n_{Z'} \langle \sigma v^2 \rangle \\
&=& n_{X} n_{Z'} \frac{\Delta_3 \alpha^3_X}{m^{5}_X},\nonumber
\end{eqnarray}
where $\Delta_2$ and $\Delta_3$ are order one coefficients. The rate for the later process (which dominates over the former process for $T_h \lsim m_X$) exceeds the Hubble rate (in the $T_h \gg m_{Z'}$ limit) when:
\begin{eqnarray}
g_X \bigg(\frac{m_X T_h}{2\pi}\bigg)^{3/2} \exp\bigg(\frac{-m_X}{T_h} \bigg)  \, \bigg(\frac{\zeta(3)g_{Z'} T^3_h}{\pi^2}\bigg)  \, \frac{\Delta_3 \alpha^3_X}{m^{5}_X} \nonumber \\
\gsim \bigg(\frac{4\pi^3 g^{\rm eff}_{\star} T^4_h}{45 m^2_{\rm Pl} \xi^4}\bigg)^{1/2}, \nonumber \\
\end{eqnarray}
which is satisfied in the parameter range of interest for $T_h \gsim m_X/10$. These processes are therefore capable of maintaining chemical equilibrium among the $Z'$ population until the temperature drops below $T_h \sim m_X/10$. At that point, the comoving $Z'$ number density becomes fixed, until they ultimately decay.


We note that in our numerical results presented in Figs.~\ref{example} and~\ref{mainfig}, we have taken the abundance of the $Z'$ population to be similar to the equilibrium value throughout the process of $X$ freeze-out. It is possible, however, that the processes capable of changing the total number of hidden sector particles (such as those described above) may become inefficient prior to the freeze-out of $X$. In this case, the $Z'$ population will depart from chemical equilibrium, $n_{Z'} \ne n^{\rm eq}_{Z'}$, altering the relic abundance, $\Omega_X h^2$. 

To assess the error that this approximation introduces, we have compared our results to the numerical solution to the coupled system of Boltzmann equations for $X$ and $Z'$. For $m_X /m_{Z'} \gsim 10$, as considered in this letter, we find that the value of $\Omega_X h^2$ is impacted at only the $\mathcal{O}(10)\%$ level. 

\subsection{Freeze-Out Temperature}
\label{fotemp}

$x_f$ is defined as the ratio of the mass of the DM particle to the visible sector temperature at freeze-out. This quantity is found by solving the following equation by iteration:
\begin{eqnarray}
\hspace{-0.2cm}\frac{x_f}{\xi}  \approx  \ln \! \bigg[\frac{c(c+2)}{4\pi^3} \! \bigg(\frac{45 \xi^5}{2g^{\rm eff}_{\star} x_f}\bigg)^{\!\!1/2}  \!\frac{g_X m_X m_{\rm Pl} (a+6\xi b/x_f)}{(1-3\xi/2 x_f)} \bigg].~~~~~~~~ \nonumber \\
\end{eqnarray}
Taking the parameter $c = 0.4$ to match numerical results, this yields:
\begin{eqnarray}
\frac{x_f}{\xi}\approx 20.8 &-& \ln\bigg(\frac{m_X}{{\rm PeV}}\bigg) + 2 \ln\bigg(\frac{\alpha_X}{0.1}\bigg) + \frac{5}{2}\ln \bigg(\frac{\xi}{1.3}\bigg)    \nonumber \\
 &-&  \frac{1}{2} \ln\bigg(\frac{x_f}{27.5}\bigg) - \frac{1}{2}\ln \bigg(\frac{g^{\rm eff}_{\star}}{100}\bigg),~~~~~~~~ 
\end{eqnarray}
which recovers the conventional WIMP expectation $x_f \sim 20$ in the $\xi = 1$ limit.

\subsection{Elastic Scattering With Nuclei}
\label{direct}

An upper limit on $\epsilon$ can be placed from the null results of direct detection experiments. The elastic scattering cross section between DM and a nucleus of atomic mass $A$ and atomic number $Z$ is given by:
\begin{eqnarray}
\sigma_{X N} &=&  \mu^2 \alpha_X \bigg[ Z \bigg(\frac{(2g_{uv}+g_{dv})}{m^2_{Z'}} + \frac{g_Y (\frac{1}{4}-\sin^2 \theta_W) \sin\theta_{Z'}}{\sin \theta_W\,m^2_Z}\bigg)  \nonumber \\
&+& (A-Z) \bigg(\frac{(g_{uv}+2g_{dv})}{m^2_{Z'}} - \frac{g_Y \sin \theta_{Z'}}{4 \sin \theta_W\,m^2_Z}\bigg) \bigg]^2, 
\end{eqnarray}
where $\mu$ is the reduced mass. Here we have included terms resulting from both $Z'$ and $Z$ exchange. In the $m_{Z'} \gg m_Z$ limit, the mixing angle between the $Z$ and $Z'$ is $\sin \theta_{Z'} \simeq -\epsilon \sin \theta_W m^2_Z/m^2_{Z'}$~\cite{Hoenig:2014dsa}, and the cross section reduces to:
\begin{eqnarray}
\sigma_{X N} &=& \frac{4 \mu^2 \alpha_X \epsilon^2 g^2_Y \cos^4 \theta_W \, Z^2}{m^4_{Z'}}. 
\end{eqnarray}

\subsection{Constraints From BBN}
\label{bbnsec}

If the decays of the $Z'$ population reheat the universe to a temperature lower than $\sim$1-10 MeV, this would likely destroy the successful predictions of Big Bang Nucleosynthesis (BBN)~\cite{Cyburt:2015mya}.  By setting the lifetime of the $Z'$ equal to $2/3H$, we find that these decays reheat the universe to a temperature given by: 
\begin{equation}
T_{\rm RH} \approx \left(\frac{5}{\pi^3 g_*}\right)^{1/4} \,   \sqrt{    m_{\rm Pl} \Gamma_{Z'}       } \, ,~~
\label{bbn}
\end{equation}
where $g_{\star}$ is the effective number of relativistic degrees-of-freedom at temperature $T_{\rm RH}$ and the width is given by: 
\begin{eqnarray}
\Gamma_{Z'}  &=& \sum_f \frac{n_c  \,  m_{Z'}   \beta_f  }{12\pi} \left[g_{fv}^2 \left(1+ \frac{ 2m^2_f}{m^2_{Z'}} \right) +g^2_{fa} \left(1-\frac{4m^2_f}{m^2_{Z'}} \right)\right],  \nonumber \\
\end{eqnarray}
where $n_c$ is the number of colors of the final state fermions and $\beta_f \equiv \sqrt{1-4m_f^2/m^2_{Z'}}$ is their velocity. For $m_{Z'} \gg m_Z, m_f$, and summing over all SM fermions, this reduces to the following lifetime:
\begin{eqnarray}
\tau_{Z'}   &\approx& 3.9 \times 10^{-8} \, {\rm s} \times \bigg(\frac{10^{-10}}{\epsilon}\bigg)^2 \bigg(\frac{100 \, {\rm TeV} }{m_{Z'}}\bigg). 
\end{eqnarray}
By requiring that the decays of the $Z'$ population do not reheat the universe to a temperature below $\sim$10 MeV, potentially destroying the successful predictions of BBN, we must restrict $\epsilon$ to the following:
\begin{eqnarray}
\epsilon \gsim 2 \times 10^{-13} \times \bigg(\frac{100 \, {\rm TeV}}{m_{Z'}}\bigg)^{1/2} \, \bigg(\frac{g_{\star}}{10}\bigg)^{1/4}.
\end{eqnarray}

\subsection{Equilibrium Between the Hidden and Visible Sectors}
\label{kesec}

Equilibrium between the hidden and visible sectors is obtained if the rate of $\gamma f \leftrightarrow Z' f$ scattering exceeds the rate of Hubble expansion:
\begin{eqnarray}
\sum_f \sigma_{\gamma f \leftrightarrow Z' f} v \,  n_f  \gsim \bigg(\frac{4 \pi^3 g^{\rm eff}_{\star} T^4}{45 m^2_{\rm Pl}}\bigg)^{1/2},
\label{condition}
\end{eqnarray}
where in the $\sqrt{s}, m_{Z'} \gg m_f$ limit:
\begin{eqnarray}
\sigma_{\gamma f \rightarrow Z' f} v &\approx& \frac{\alpha Q_f^2 (g_{fv}^2+g_{fa}^2)}{4 s^2} \bigg[s+6m^2_{Z'} -\frac{7m^4_{Z'}}{s} \\
&+& 2\bigg(s-2m^2_{Z'} +\frac{2m^4_{Z'}}{s} \bigg)\ln\bigg(\frac{s(1-m^2_{Z'}/s)^2}{m_f^2}\bigg) \bigg], \nonumber
\end{eqnarray}
and
\begin{eqnarray}
\sigma_{Z' f \rightarrow \gamma f} v &\approx& \frac{\alpha Q_f^2 (g_{fv}^2+g_{fa}^2)}{6 (s-m^2_{Z'})^2} \bigg[s+6m^2_{Z'} -\frac{7m^4_{Z'}}{s} \\
&+& 2\bigg(s-2m^2_{Z'} +\frac{2m^4_{Z'}}{s} \bigg)\ln\bigg(\frac{s(1-m^2_{Z'}/s)^2}{m_f^2}\bigg) \bigg]. \nonumber
\end{eqnarray}
Combining this with $n_f = 3 \zeta(3) g_f T^3/4 \pi^2$, and approximating $\sqrt{s} \approx 4 \, T$, we find that Eq.~\ref{condition} is satisfied if $\epsilon \gsim  10^{-7} \times (T/10\,{\rm GeV})^{1/2} \, (g_{\star}/100)^{1/4}$.

\end{appendix}

\bibliography{earlyhiddenletter}

\end{document}